\documentclass[rnote]{aa} 
\usepackage{graphicx}
\usepackage{txfonts}
\begin{document}

   \title{Calibrating the projection factor for Galactic Cepheids}


   \author{Chow-Choong Ngeow\inst{1}
          \and
          Hilding R. Neilson\inst{2}
          \and
          Nicolas Nardetto\inst{3}
          \and
          Massimo Marengo\inst{4}
          }

   \institute{Graduate Institute of Astronomy, National Central University, Jhongli City, 32001, Taiwan\\
              \email{cngeow@astro.ncu.edu.tw}
         \and
         Argelander Institute for Astronomy, Auf dem Huegel 71, 53121 Bonn, Germany
         \and
         Laboratoire Lagrange, UMR7293, UNSA/CNRS/OCA, 06300 Nice, France 
         \and
         Department of Physics and Astronomy, Iowa State University, Ames, IA 50010, USA 
   }

   \date{Received 06 January, 2012; accepted 18 May, 2012}

   
 
  \abstract
   {The projection factor ($p$), which converts the radial velocity to pulsational velocity, is an important parameter in the Baade-Wesselink (BW) type analysis and distance scale work. The $p$-factor is either adopted as a constant or linearly depending on the logarithmic of pulsating periods.}
   {The aim of this work is to calibrate the $p$-factor if a Cepheid has both the BW distance and an independent distance measurement, and examine the $p$-factor for $\delta$ Cephei -- the prototype of classical Cepheids.}
   {We calibrated the $p$-factor for several Galactic Cepheids that have both the latest BW distances and independent distances either from Hipparcos parallaxes or main-sequence fitting distances to Cepheid-hosted stellar clusters.}
   {Based on $25$ Cepheids, the calibrated $p$-factor relation is consistent with latest $p$-factor relation in literature. The calibrated $p$-factor relation also indicates that this relation may not be linear and may exhibit an intrinsic scatter. We also examined the discrepancy of empirical $p$-factors for $\delta$ Cephei, and found that the reasons for this discrepancy include the disagreement of angular diameters, the treatment of radial velocity data, and the phase interval adopted during the fitting procedure. Finally, we investigated the impact of the input $p$-factor in two BW methodologies for $\delta$ Cephei, and found that different $p$-factors can be adopted in these BW methodologies and yet result in the same angular diameters.}
   {}


   \maketitle
%

\section{Introduction}

The projection factor, or $p$-factor, converts the observed radial velocity to pulsational velocity and is a key parameter in the Baade-Wesselink (BW) type analysis and in distance scale applications. Currently the $p$-factor in the literature is either adopted as a constant with the period (some of the values are given, for example, in Burki et al. \cite{bur82}; Kov{\'a}cs \cite{kov03}; Groenewegen \cite{gro07}; Feast et al. \cite{fea08}) or involves a period-dependence in the form of $p=a\log(P)+b$ (the $Pp$ relation, see, for example, Gieren et al. \cite{gie93,gie05}; Nardetto et al. \cite{nar07,nar09}; Laney \& Joner \cite{lan09}; Storm et al. \cite{sto11a}). These $p$-factors are listed in Table \ref{tab_pfactor}. The latest derivation of a $Pp$ relation by Storm et al. (\cite{sto11a}, hereafter S11), is based on the combination of two constraints: Galactic Cepheids possessing accurate parallaxes from the {\it Hubble Space Telescope} ({\it HST}, Benedict et al. \cite{ben02}, \cite{ben07}) to constrain the intercept of the $Pp$ relation, and Large Magellanic Cloud (LMC) Cepheids to determine the $Pp$ slope by demanding that BW distances to these LMC Cepheids be independent of the pulsation period. The $p$-factor is found to be insensitive to variations in chemical abundance (Nardetto et al. \cite{nar11}).

\begin{table}
\caption{Comparison of the $p$-factor.}            
\label{tab_pfactor}   
\centering            
\begin{tabular}{lcc}   
\hline\hline           
Reference & $a$ & $b$  \\  
\hline                       
Burki et al. (\cite{bur82})    & $\cdots$         & $1.36$ \\
Kov{\'a}cs (\cite{kov03})      & $\cdots$         & $1.35$ \\
Groenewegen (\cite{gro07})     & $\cdots$         & $1.27\pm0.05$ \\
Feast et al. (\cite{fea08})    & $\cdots$         & $1.23\pm0.03$ \\
Gieren et al. (\cite{gie93})   & $-0.03$          & $1.39$ \\
Gieren et al. (\cite{gie05})   & $-0.15\pm0.02$   & $1.58\pm0.02$ \\
Nardetto et al. (\cite{nar07}) & $-0.075\pm0.031$ & $1.366\pm0.036$ \\
Nardetto et al. (\cite{nar09}) & $-0.08\pm0.05$   & $1.31\pm0.06$ \\
Laney \& Joner (\cite{lan09})  & $-0.071\pm0.020$ & $1.311\pm0.019$ \\
Storm et al. (\cite{sto11a})   & $-0.186\pm0.06$  & $1.550\pm0.04$ \\
\hline                               
\end{tabular}\\
\end{table}

The {\it Gaia} mission will not present distances to Galactic Cepheids in the next few years (though it is scheduled to launch in 2013). Until then, the BW method is the best avenue for measuring the Cepheid parallax beyond {\it HST} and main-sequence fitting, but the method depends on the $p$-factor, which is known to depend on atmospheric physics, the circumstellar medium, and observational bias (e.g., the way radial velocities are measured). The goal of this paper is to extend the work of S11. Because S11's $Pp$ relation was calibrated using Cepheids with {\it HST} parallaxes (and LMC Cepheids), we demonstrate in Section 2 that the $Pp$ relation can also be calibrated using other independent distances. The discrepancy of the $p$-factor for $\delta$ Cephei is discussed in Section 3, we explore in which way different methods affect the $p$-factor determination by using different angular diameter measurements and BW methods. The conclusion is given in Section 4.

\section{Calibration of the $Pp$ relation using independent distances}

Because the $p$-factor is degenerate with the measured distance for a given Cepheid (for example, see Barnes et al. \cite{bar05a}), the $p$-factor can be calibrated if a Cepheid has both the BW distance (with an adopted $p$-factor) and an independent distance measured from other methods. That is

\begin{figure}
  \resizebox{\hsize}{!}{\includegraphics{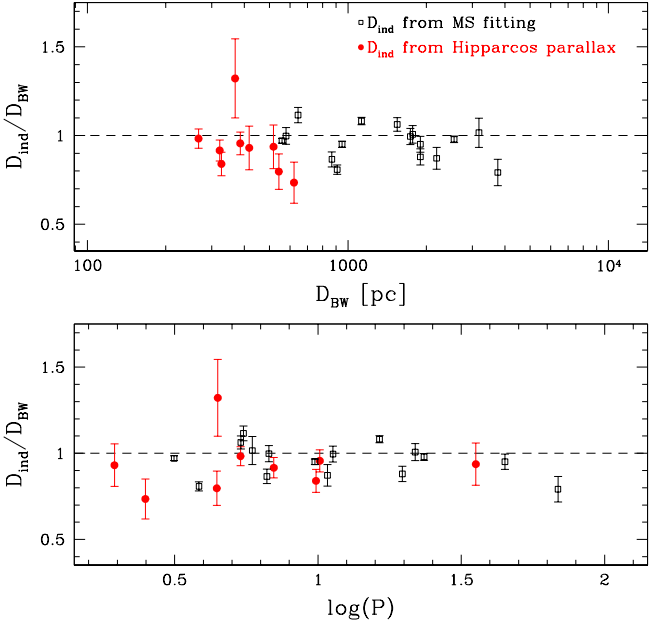}}
  \caption{{\bf Top panel (a):} Distance ratio as a function of BW distances given in S11. {\bf Bottom panel (b):} Distance ratio as a function of pulsation periods in days. In both panels, the dashed lines indicate the case for $D_{\mathrm{ind}}/D_{\mathrm{BW}}=1$ and not the fit to the data. [See on-line edition for a color version.]}
  \label{fig1}
\end{figure}

\begin{eqnarray}
p_{\mathrm{new}} & = & p_{\mathrm{BW}} \times \frac{D_{\mathrm{ind}}}{D_{\mathrm{BW}}}, 
\end{eqnarray} 

\noindent where $D$ is the distance in parsec. This was performed, for example, in Groenewegen (\cite{gro07}), Feast et al. (\cite{fea08}), and Laney \& Joner (\cite{lan09}), who compared the BW distances and geometric parallaxes to calibrate the $p$-factor. Because {\it HST} parallaxes have been used to constraint the $Pp$ relations in S11, the derived BW distances are not fully independent of these parallaxes. Therefore, we adopted the revised {\it Hipparcos} parallaxes (van Leeuwen et al. \cite{van07}; their Tables 1 and 2) for nine Cepheids\footnote{We excluded Y Sgr for the reason given in van Leeuwen et al. (\cite{van07}), and RT Aur because it has a negative parallax.}, augmented with distances based on the main-sequence (MS) fitting technique from Turner (\cite{tur10}) for another $16$ Cepheids, common to the S11 sample\footnote{W Sgr was excluded from the sample based on the reasons given in S11.} to calibrate the $p$-factors. Note that the MS distance for TW Nor was updated based on the latest result from Majaess et al. (\cite{maj11a}). The averaged distance ratio for these $25$ Cepheids is $<D_{\mathrm{ind}}/D_{\mathrm{BW}}>=0.966$, with a dispersion ($\sigma$) of $0.123$. Figure \ref{fig1} shows the distribution of $D_{\mathrm{ind}}/D_{\mathrm{BW}}$ as a function of BW distances (upper panel) and pulsation periods (lower panel). A weak period dependence is found for the distance ratio: $D_{\mathrm{ind}}/D_{\mathrm{BW}}=-0.019(\pm0.065)\log P+0.969(\pm0.067)$ with $\sigma=0.124$, which is consistent with being period-independent.

\begin{figure}
  \resizebox{\hsize}{!}{\includegraphics{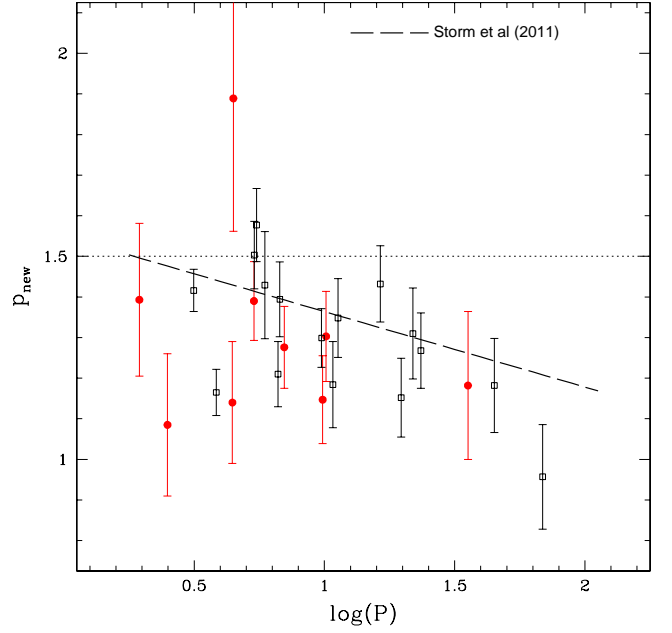}}
  \caption{Calibrated $p$-factor for individual Cepheids using Eq. (1). The dashed line indicates the expected limit of the $p$-factor ($p=1.5$, see text for more details). The $Pp$ relation from S11 is included for comparison. Symbols are the same as in Fig. \ref{fig1}. [See on-line edition for a color version.]}
  \label{fig2}
\end{figure}

\begin{figure}
  \resizebox{\hsize}{!}{\includegraphics{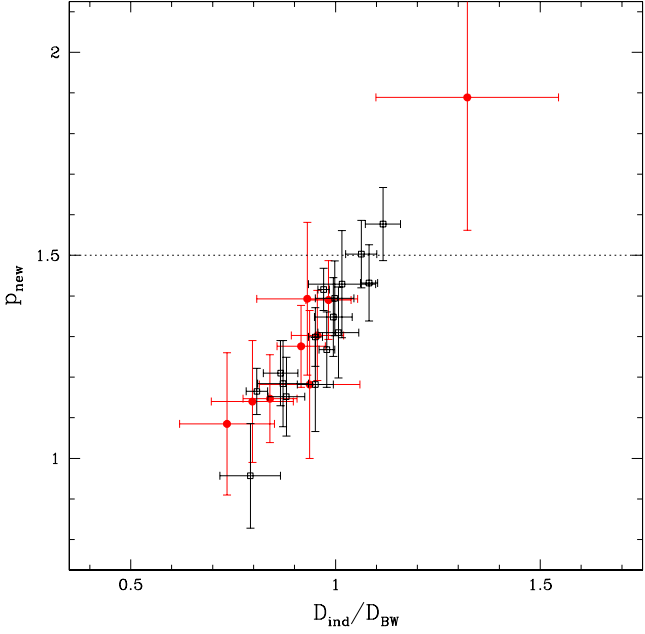}}
  \caption{Correlation of distance ratios and calibrated $p$-factors for Cepheids in the sample. The dashed line indicates the expected limit of $p$-factor ($p=1.5$, see text for more details). The slope of the data at the point where the intercept passes through the origin, is $1.374$, which is almost identical to $-0.186\ \times <\log (P)> + 1.550=1.372$, with a mean period of $<\log (P)>=0.955$. Symbols are the same as in Fig. \ref{fig1}. The outlier with largest error bars is FF Aql. [See on-line edition for a color version.]}
  \label{fig3}
\end{figure}

Distance ratios for individual Cepheids in the sample were used to calibrate the $p$-factors using Eq. (1). The resulting $p$-factors as a function of pulsation periods are presented in Fig. \ref{fig2}, where the error bars include propagated errors from both distances and the $Pp$ relation from S11. A linear regression fit to all data yields $p=-0.172(\pm0.086)\log P+1.462(\pm0.087)$, with $\sigma=0.107$, and it is consistent with S11's $Pp$ relation. Figure \ref{fig3} presents the correlation of the calibrated $p$-factor, $p_{\mathrm{new}}$, as a function of the distance ratios. Figures \ref{fig2} and \ref{fig3} reveal that about three Cepheids have calibrated $p$ values ($p_{\mathrm{new}}$) that fall outside the expected limit of the $p$-factor: $p>1.5$. These values indicate that limb brightening instead of limb darkening occurs in a Cepheid atmosphere (S11). This can be seen from Eq. (6) in Nardetto et al. (\cite{nar06}): $p_c=-0.18u_V+1.52$, where $p_c$ is geometric $p$-factor and $u_V$ is the limb darkening in $V$ band. For a uniform limb darkening, $u_V=0$ when $p\sim1.5$, and hence $p>1.5$ implies a limb brightening (also, see Neilson et al. \cite{nei11a}). Since $p_{\mathrm{new}}$ is degenerate with $D_{\mathrm{ind}}/D_{\mathrm{BW}}$, as shown from Eq. (1), $p_{\mathrm{new}}>1.5$ for these Cepheids suggested that either the independent distances are overestimated or the BW distances are underestimated, or both distances are incorrect. Nevertheless, these Cepheids have $p_{\mathrm{new}}$ within $\sim \sigma_p$ from the limit (where $\sigma_p$ is the estimated error on $p_{\mathrm{new}}$). The discrepant point shown in Fig. \ref{fig2} is FF Aql, with $p=1.89$. This is due to the large difference in distances from {\it HST} ($\pi_{HST}=2.81\pm0.18$~mas [milli-arcsecond], or $D=355.9\pm22.8$~pc) and {\it Hipparcos} ($\pi_{\mathrm{hipparcos}}=2.05\pm0.34$~mas, or $D=487.8\pm80.9$~pc). After removing FF Aql, the resulting $Pp$ relation is: $p=-0.159(\pm0.070)\log P+1.447(\pm0.070)$, with a dispersion of $0.064$. The $Pp$ relation from S11 would be a preferred relation, because it is calibrated with {\it HST} parallaxes and constrained from LMC Cepheids.

Based on a slightly larger sample of Galactic Cepheids than that of S11, Fig. \ref{fig2} reveals that the $p$-factor relation may not be a constant\footnote{For completeness, the weighted mean for the $p$-factors shown in Fig. \ref{fig2} is $1.306\pm0.019$ ($\sigma=0.187$), or $1.304\pm0.019$ ($\sigma=0.148$) if FF Aql is removed.}. Furthermore, the period dependency of $Pp$ relation may not be linear, especially if the period-luminosity (P-L) relation is nonlinear (see, for example, Kanbur \& Ngeow \cite{kan04}; Sandage et al. \cite{san04}; Ngeow et al. \cite{nge05}, \cite{nge09}; Neilson et al. \cite{nei10}, and reference therein). From the perspective of the geometric $p$-factor, it depends on limb darkening, which in turn depends on the combination of luminosity, effective temperature, and gravity (Neilson \& Lester \cite{nei11}). If the P-L and period-temperature (or color) relations are nonlinear, then the same holds for the geometric $Pp$ relation. This nonlinearity could be in quadratic or in other forms\footnote{For examples, two linear regressions with a break period at 10 days, or in the form of a constant $+$ linear regressions, or a power law relation.}. Nevertheless, confirmation or refutation of nonlinearity of the $p$-factor relation has to wait for accurate parallax measurements from {\it Gaia} mission for Cepheids in S11 sample. Furthermore, based on spherically symmetric atmosphere models, Neilson et al. (\cite{nei11a}) found that the theoretical $Pp$ relation is also nonlinear. 

Figure \ref{fig2} also suggests a possible existence of intrinsic dispersion on the $Pp$ relation, albeit large error bars, which could be caused by systematics in the BW or independent distances, or both. Indeed, the $p$-factor measured in this work includes all potential uncertainties in the BW method (or there is a problem in the implementation of the BW method). The dispersion may be a natural result of the width of the instability strip, i.e. period-dependent relations for Cepheids typically exhibit an intrinsic dispersion (e.g., the period-color relation). Another possible source for the dispersion may be dynamics in Cepheid atmospheres. Assuming that pulsation amplitudes provide a measure of atmospheric dynamics, then the significant dispersion in period-amplitude relations (as shown in Klagyivik \& Szabados \cite{kla09}) suggests that dynamics can contribute to the significant dispersion of the $Pp$ relation. Again, testing the existence of intrinsic dispersion has to await parallax measurements from {\it Gaia} mission.

\section{The $p$-factor for $\delta$ Cephei}

\begin{figure}
  \resizebox{\hsize}{!}{\includegraphics{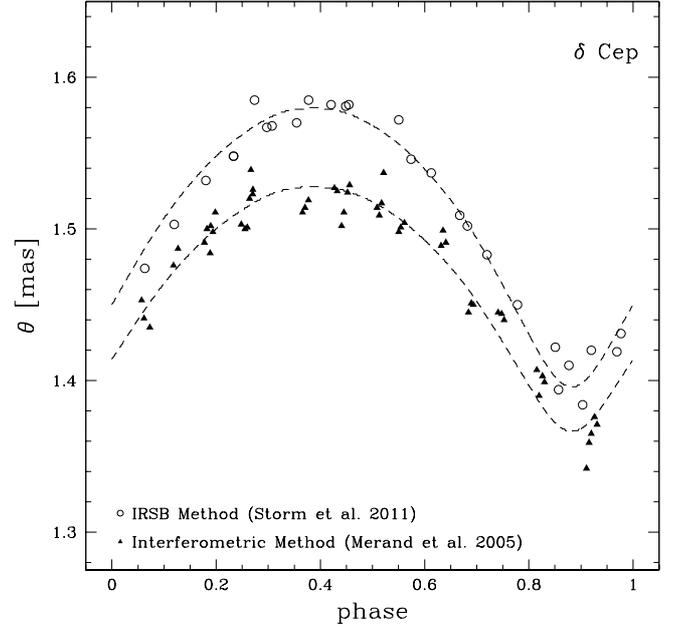}}
  \caption{Comparison of the angular diameters for $\delta$ Cephei as a function of pulsation phases from two methods. These two methods are the interferometric technique (M{\'e}rand et al. \cite{mer05}, their Table 3) and IRSB technique using the prescription given in S11, with $VK$ band photometric light curve data taken from Moffett \& Barnes (\cite{mof84}) and Barnes et al. (\cite{bar97}). Epoch of $2443674.144$ days and period of $5.366316$ days are adopted from Moffett \& Barnes (\cite{mof85}). Error bars are omitted for clarity. The dashed curves are modeled angular diameters as a function of pulsational phases based on the solution of Eq. (2), using $D=273.2$~pc, see text for more details.}
  \label{fig_deltacep}
\end{figure}

M{\'e}rand et al. (\cite{mer05}) have directly measured the $p$-factor, $p=1.27\pm0.06$, for $\delta$ Cephei using interferometric measurements and the {\it HST} parallax. This value is consistent with the result derived from theoretical predictions ($p=1.27\pm0.01$, Nardetto et al. \cite{nar04}) or with the value of $p=1.24\pm0.05$ from the $Pp$ relation provided in Nardetto et al. (\cite{nar09}). The $p$-factor established by M{\'e}rand et al. (\cite{mer05}) and Nardetto et al. (\cite{nar09}) is tied to radial velocities inferred from the cross-correlation method, whereas Nardetto et al. (\cite{nar04}) relied on applying Gaussian fits to determine the radial velocities. In contrast, the $p$-factor derived from using the BW distance and {\it HST} parallax is $1.45\pm0.07$ (S11), about $2\sigma$ larger than the value given in M{\'e}rand et al. (\cite{mer05}). The predicted $p$-factor using the $Pp$ relation from S11 is $1.41$, which is still larger than the value from M{\'e}rand et al. (\cite{mer05}). In this section, we examine the possible reasons for this discrepancy.

\subsection{Angular diameters for $\delta$ Cephei}

Figure \ref{fig_deltacep} compares the computed angular diameters as established from interferometric technique (M{\'e}rand et al. \cite{mer05}) and infrared surface brightness (IRSB) technique as a function of pulsation phases. It is clear from this figure that the angular diameters from these two techniques do not agree, especially at phases from $\sim0.2$ to $\sim0.7$, even though Kervella et al. (\cite{ker04}) have shown a good agreement for $\ell$ Car. The mean angular diameters from interferometric and IRSB techniques are $1.475$~mas (with $\sigma=0.053$) and $1.508$~mas (with $\sigma=0.068$), respectively. Cepheid angular diameters measured from interferometric observations have been corrected using limb-darkening relations from model stellar atmospheres. Marengo et al. (\cite{mar03}) found that the model corrections introduced negligible errors and were largely insensitive to atmospheric dynamics. That analysis assumed plane-parallel radiative transfer, but Neilson \& Lester (\cite{nei11}) found that for stars with low gravity ($\log g \sim 1$ - $3$), more realistic spherically symmetric model atmospheres predicted greater variations of limb darkening than those found by Marengo et al. (\cite{mar03}). Neilson et al. (\cite{nei11a}) also showed that angular diameter corrections from plane-parallel model atmosphere lead to an approximately 2\% underestimate of the angular diameter, which gives a $\sim2$\% systematic underestimate of the $p$-factor. It is likely that the interferometric observations predict angular diameters that are smaller than predicted by the IRSB, which measured the angular diameter from the stellar flux and temperature-color relations. Nevertheless, Nardetto et al. (\cite{nar06b}) showed that the derived distance to $\delta$ Cephei is not affected by the limb-darkening variation with phases on the interferometric angular diameters. 

Another possibility is that the angular diameters from the IRSB method is overestimated. Neilson et al. (\cite{nei10}) found that angular diameters measured using the IRSB techniques may be overestimated because of circumstellar media that cause an infrared (IR) excess. This IR excess is presumably caused by mass-loss activity, evidence for which has been found recently on $\delta$ Cephei (Marengo et al. \cite{mar10b}; Matthews et al. \cite{mat12}). In particular, the $K$-band flux excess for $\delta$ Cephei is 1.5\% (M{\'e}rand et al. \cite{mer06}), which translates into a difference of $\Delta K=0.016$ magnitudes. Using the surface brightness relation (S11), this flux excess translates into a difference in angular diameter, $\Delta \log \theta = 0.004$. For a mean angular diameter of $\sim1.48$~mas, the IRSB technique would overestimate the angular diameter by $\Delta \theta =0.01$ -- $0.02$~mas or $\Delta \theta/\theta = 1\%$. $K$-band flux excess may explain a significant fraction of the difference seen in Fig. \ref{fig_deltacep}. However, a $1\%$ difference in angular diameter leads to a $1\%$ difference in $p$-factor, suggesting the $K$-band flux excess is insufficient to explain the difference in $p$-factor for $\delta$ Cephei found by M{\'e}rand et al. (\cite{mer05}) and S11. 

\subsection{The $p$-factor derived from the BW method}

\begin{figure}
  \resizebox{\hsize}{!}{\includegraphics{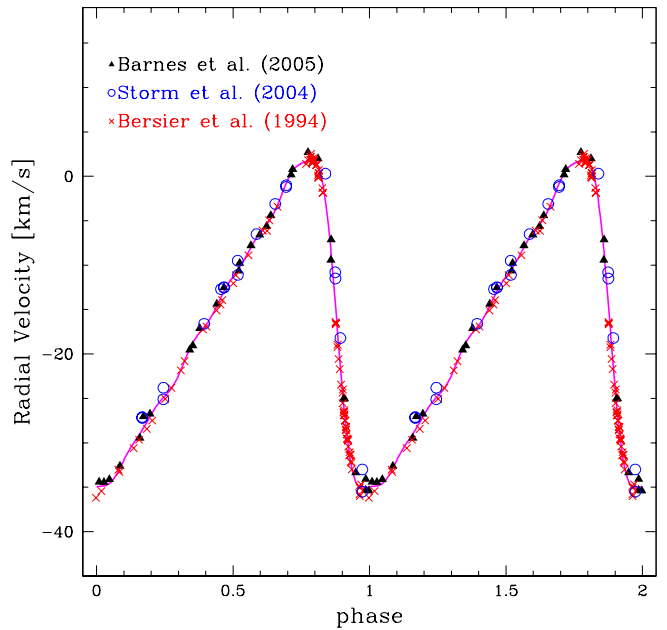}}
  \caption{Folded radial velocities for $\delta$ Cephei, using an epoch of $2443674.144$ days and period of $5.366316$ days. The data were fitted with an $8^{\mathrm{th}}$ order Fourier expansion, shown by the solid curve. Data from Bersier et al. (\cite{ber94}) needed to be shifted slightly to minimize the scatter in the combined radial velocity curve. Error bars are omitted for clarity. [See on-line edition for a color version.]}
  \label{fig_rv}
\end{figure}

In the BW analysis, the angular diameters as shown in Fig. \ref{fig_deltacep} can be modeled according to the following equation:

\begin{eqnarray}
\theta(\phi) & = & \theta_0 - p\frac{2PC}{D}\int^{\phi}_{0} [V_R(\phi)-\gamma]d\phi,
\end{eqnarray} 

\noindent where $\phi$ is the pulsational phase, $V_R$ is the radial velocity curve, $\gamma$ is the systemic velocity, and $C=0.57749$ is the conversion factor (for $P$ in days, $D$ in parsec, velocities in $\mathrm{km/s}$ and $\theta$ in~mas). Given an adopted $p$-factor, Eq. (2) can be used to derive the distance (and radius) to a Cepheid. Conversely, the $p$-factor can be determined if the distance is known a priori. The radial velocity data for $\delta$ Cephei are taken from Bersier et al. (\cite{ber94}), Storm et al. (\cite{sto04}), and Barnes et al. (\cite{bar05b}). The combined radial velocity curve is fitted with an $8^{\mathrm{th}}$ order Fourier expansion, as presented in Fig. \ref{fig_rv}, and a distance of $D=273.2$~pc was adopted for $\delta$ Cephei (Benedict et al. \cite{ben02}). The fitting procedure was performed using an OSL bi-sector algorithm from {\tt SLOPE} (Isobe et al. \cite{iso90}). The fitted $p$-factors, based on M{\'e}rand et al. (\cite{mer05}) and S11 angular diameters, are summarized in Table \ref{tab_dcep}. The solutions ($\theta_0$ and $p$) were subsequently used to construct the angular diameter curves using Eq. (2). These curves are shown in Fig. \ref{fig_deltacep}, and agree well with the empirical angular diameters. 

\begin{table}
\caption{Derived $p$-factors for $\delta$ Cep.}            
\label{tab_dcep}   
\centering            
\begin{tabular}{lcc}   
\hline\hline           
Data source for $\theta$ & Include all phases & Exclude phases $>0.8$  \\  
\hline                       
M{\'e}rand et al. (\cite{mer05}) & $1.32\pm0.05$  & $1.26\pm0.08$ \\
Storm et al. (\cite{sto11a})     & $1.50\pm0.05$  & $1.58\pm0.08$ \\
\hline                               
\end{tabular}\\
\end{table}

The derived $p$-factors using all data points, $p=1.32\pm0.05$ and $1.50\pm0.05$, are consistent with $p=1.27$ from M{\'e}rand et al. (\cite{mer05}) and $p=1.45$ from S11, respectively, although the derived values are higher by $0.05$. The $p$-factors from using the M{\'e}rand et al. (\cite{mer05}) and S11 angular diameter data disagree at $\sim2.5\sigma$ level, due to the disagreement of the angular diameters shown in Fig. \ref{fig_deltacep}. The difference of the $p$-factor stems from the manner in which the radial velocities were treated, since a similar (or the same) period and distance for $\delta$ Cephei were adopted in M{\'e}rand et al. (\cite{mer05}), S11, and this work. M{\'e}rand et al. (\cite{mer05}) considered radial velocity data either from Bersier et al. (\cite{ber94}) or Barnes et al. (\cite{bar05b}); while S11 included several additional data sources to those adopted in this work. For fitting the radial velocity data, M{\'e}rand et al. (\cite{mer05}) applied a four-knot periodic cubic spline interpolation, in contrast to S11 and this work, where a Fourier expansion was used to fit the radial velocity data\footnote{The difference of $p$-factors from S11 and Table \ref{tab_dcep} may include the different order of the Fourier expansion: S11 adopted a lower order fit while in this work higher order terms are included.}. 

Finally, M{\'e}rand et al. (\cite{mer05}) used all data points during the fitting procedure, while S11 excluded data points with $\phi>0.8$ due to the deviation of angular diameters at these phases during the fitting (probably caused by shock waves at minimum radius, see also Fouqu{\'e} et al. \cite{fou03};  Kervella et al. \cite{ker04}). In Table \ref{tab_dcep}, we also include the fitted $p$-factors if data points for $\phi>0.8$ are excluded, as in S11. These $p$-factors agree with the $p$-factors derived from using all data, but they disagree at $\sim2.8\sigma$ level (the reason is again owing to the disagreement of the angular diameters). Nevertheless, it is clear from Table \ref{tab_dcep} that the adopted phase intervals will affect the fitted $p$-factor. 

\subsection{$p$-factor from other BW methodologies}

Feast et al. (\cite{fea08}) and Laney \& Joner (\cite{lan09}) employed a different BW method to derive the distance to $\delta$ Cephei. Their method is described in more detail in Laney \& Stobie (\cite{lan95}, hereafter LS95). After deriving the BW distance, the $p$-factor is recalibrated using geometrical parallaxes. We focus on the recalibrated $p$-factor from Laney \& Joner (\cite{lan09}), because they adopted the same {\it HST} parallax as in S11, which yields $1.289\pm0.061$. Their $p$-factor is also $\sim2\sigma$ smaller than the $p$-factor from S11. On the other hand, the BW distance of $268.8\pm7$~pc from Laney \& Joner (\cite{lan09}) is in excellent agreement with the BW distance of $266.7\pm5$~pc from S11, even though the adopted $p$-factor is different: $1.27$ versus $1.41$. This reflects that the difference in the adopted $p$-factor may compensate for the difference in assumptions and methodologies in these two BW methods.

This can also be seen from the following. In S11, the photometric surface brightness relation can be expressed as $F_V = \alpha (V-K)_0 + \beta$ and $F_V = 4.2207 - 0.1V_0 - 0.5\log \theta(\phi)$. Equating these two expressions and re-arranging them, yields

\begin{eqnarray}
V_0 & = & \alpha' (V-K)_0 -5 \log \theta(\phi) + \beta',
\end{eqnarray}

\noindent where $\alpha'=-10\alpha$ and $\beta'=10(4.2207-\beta)$ are constants, with $\alpha=-0.1336$ and $\beta=3.9530$ as given in S11. In the LS95 methodology, the radius in $V_0=a(V-K)_0-5\log (R+\Delta R) + b$ can be converted into angular diameters:

\begin{eqnarray}
V_0 & = & a (V-K)_0 -5 \log \theta(\phi) + b',
\end{eqnarray}

\begin{figure}
  \resizebox{\hsize}{!}{\includegraphics{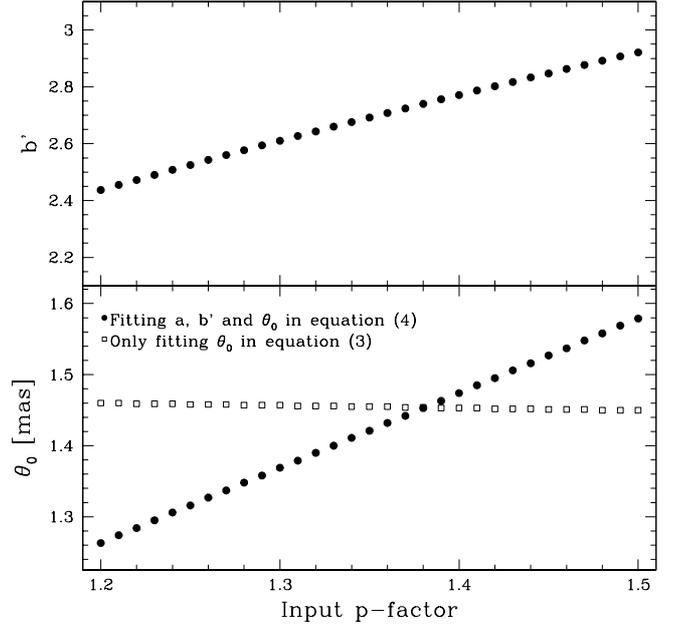}}
  \caption{Fitted values of $b'$ (upper panel) and $\theta_0$ (lower panel) in Eq. (4) as a function of input $p$-factors. Filled circles are the fitted results using Eq. (4), and open squares are the fitted $\theta_0$ using Eq. (3). Note that at $273.2$~pc, various published period-radius relations give $\theta_0$ in the range from $\sim1.4$~mas to $\sim1.7$~mas for $\delta$ Cephei.}
  \label{fig_bp}
\end{figure}

\noindent where $b'=b-5\log (D/2)$ (here, $D$ is treated as a nuisance parameter). In both Eq. (3) \& (4), the $\theta (\phi)$ is given in Eq. (2). In contrast to the S11 methodology, the coefficient $a$ and $b'$, as well as $\theta_0$, in Eq. (4) are solved (or fitted) and not fixed as constants.

Using the same $VK$ photometric data as in Fig. \ref{fig_deltacep} and the radial velocity curves as presented in Fig. \ref{fig_rv}, we fit coefficients $a$, $b'$ and $\theta_0$ in Eq. (4) for a range of input $p$-factors (from $1.20$ to $1.50$ with a step size of $0.01$). The results are presented in Fig. \ref{fig_bp} for $b'$ and $\theta_0$, the fitted value for $a$ is $1.292$ regardless of input $p$-factors (in contrast to $1.336$ given in Eq. [3]). The fitted coefficients give the same sum of residuals squared for all input $p$-factors. For comparison, we also fit the $\theta_0$ in Eq. (3) for the same input $p$-factors. The fitted results are shown as open squares in the lower panel of Fig. \ref{fig_bp}. This shows that in the LS95 methodology, $\theta_0$ is more sensitive to the input $p$-factor than the S11 methodology. Note that if the mean radius of $\delta$ Cephei is $41.9R_\odot$ (Feast et al. \cite{fea08}), then $\theta_0=1.45$~mas at the distance found in Laney \& Joner (\cite{lan09}, $D=268.8$~pc). Based on the lower panel of Fig. \ref{fig_bp}, the required input $p$-factor to obtain $\theta_0=1.45$~mas is $1.38$ and $1.48$ for the LS95 and S11 methodology, respectively. Even though the $p$-factors found here are higher than $1.27$ and $1.41$, this example demonstrates that different $p$-factors adopted in the LS95 and S11 BW methodology result in the same angular diameter. Hence, it is possible to obtain similar distances with different input $p$-factors in different BW methods by compensating for the fitted coefficients.

\section{Conclusion}

We have calibrated the $p$-factors for Galactic Cepheids using the simple fact that the $p$-factor is degenerate with distance, by using Cepheids with both BW distance and an independent distance measurement. The latest BW distances for Galactic Cepheids are provided in S11, and the independent distances were either adopted from {\it Hipparcos} parallax measurements or from the MS fitting. We also compared the $p$-factors established for $\delta$ Cephei by M{\'e}rand et al. (\cite{mer05}, interferometric technique) and S11 (IRSB). Several physical aspects might explain why the aforementioned groups disagree concerning the $p$-factors: (a) inconsistent angular diameters as inferred from the interferometric and IRSB techniques; (b) a difference in the adopted radial velocity data and the interpolating/fitting of these data; and (c) a difference in the sample selected to fit the data (i.e. excluding data with $\phi>0.8$). We also examined a different BW methodology than S11, as presented in LS95 and Feast et al. (\cite{fea08}), which adopted a different $p$-factor and yet obtained almost the same distance to $\delta$ Cephei as in S11. This is because the surface brightness coefficients are fitted from data in LS95 methodology, which can compensate for a different input $p$-factor.

Even though there are currently only ten Cepheids with accurate geometrical distances measured from {\it HST}, a larger number of Cepheids will have better parallax measurements after the launch of {\it Gaia}. These parallaxes can be used to verify the $Pp$ relation derived in S11, and to examine the nonlinearity and possible intrinsic dispersion of the relation. Furthermore, direct $p$-factor measurements are being obtained for a few additional Cepheids using interferometric techniques (Kervella 2011 --- private communication), and the results will be employed to evaluate the viability of the conclusions presented here.

\begin{acknowledgements}

We thank the referee, together with D. Majaess, P. Fouqu{\'e}, P. Kervella \& A. M{\'e}rand, for useful comments to improve this manuscript. CCN thanks the funding from National Science Council (of Taiwan) under the contract NSC 98-2112-M-008-013-MY3. HRN acknowledges funding from the Alexander von Humboldt Foundation.     

\end{acknowledgements}


\begin{thebibliography}{}

\bibitem[1997]{bar97} Barnes, T.~G., III, Fernley, J.~A., Frueh, M.~L., et al.\ 1997, \pasp, 109, 645 

\bibitem[2005a]{bar05a} Barnes, T.~G., III, Storm, J., Jefferys, W.~H., Gieren, W.~P., \& Fouqu{\'e}, P.\ 2005a, \apj, 631, 572 

\bibitem[2005b]{bar05b} Barnes, T.~G., III, Jeffery, E.~J., Montemayor, T.~J., \& Skillen, I.\ 2005b, \apjs, 156, 227 

\bibitem[2002]{ben02} Benedict, G.~F., McArthur, B.~E., Fredrick, L.~W., et al.\ 2002, \aj, 124, 1695 

\bibitem[2007]{ben07} Benedict, G.~F., McArthur, B.~E., Feast, M.~W., et al.\ 2007, \aj, 133, 1810

\bibitem[1994]{ber94} Bersier, D., Burki, G., Mayor, M., \& Duquennoy, A.\ 1994, \aaps, 108, 25 

\bibitem[1982]{bur82} Burki, G., Mayor, M., \& Benz, W.\ 1982, \aap, 109, 258

\bibitem[2008]{fea08} Feast, M.~W., Laney, C.~D., Kinman, T.~D., van Leeuwen, F., \& Whitelock, P.~A.\ 2008, \mnras, 386, 2115 

\bibitem[2003]{fou03} Fouqu{\'e}, P., Storm, J., \& Gieren, W.\ 2003, Stellar Candles for the Extragalactic Distance Scale, Ed. D. Alloin \& W. Gieren, Lecture Notes in Physics, Springer-Verlag, 635, 21 

\bibitem[1993]{gie93} Gieren, W.~P., Barnes, T.~G., III, \& Moffett, T.~J.\ 1993, \apj, 418, 135 

\bibitem[2005]{gie05} Gieren, W., Storm, J., Barnes, T.~G., III, Fouqu{\'e}, P., Pietrzy{\'n}ski, G., \& Kienzle, F.\ 2005, \apj, 627, 224 

\bibitem[2007]{gro07} Groenewegen, M.~A.~T.\ 2007, \aap, 474, 975

\bibitem[1990]{iso90} Isobe, T., Feigelson, E.~D., Akritas, M.~G., \& Babu, G.~J.\ 1990, \apj, 364, 104 

\bibitem[2004]{kan04} Kanbur, S.~M., \& Ngeow, C.-C.\ 2004, \mnras, 350, 962 

\bibitem[2004]{ker04} Kervella, P., Fouqu{\'e}, P., Storm, J., et al.\ 2004, \apjl, 604, L113 

\bibitem[2009]{kla09} Klagyivik, P., \& Szabados, L.\ 2009, \aap, 504, 959 

\bibitem[2003]{kov03} Kov{\'a}cs, G.\ 2003, \mnras, 342, L58

\bibitem[1995]{lan95} Laney, C.~D., \& Stobie, R.~S.\ 1995, \mnras, 274, 337 

\bibitem[2009]{lan09} Laney, C.~D., \& Joner, M.~D.\ 2009, American Institute of Physics Conference Series, 1170, 93 

\bibitem[2011]{maj11a} Majaess, D., Turner, D., Moni Bidin, C., et al.\ 2011, \apjl, 741, L27 

\bibitem[2003]{mar03} Marengo, M., Karovska, M., Sasselov, D.~D., et al.\ 2003, \apj, 589, 968 

\bibitem[2010]{mar10b} Marengo, M., Evans, N.~R., Barmby, P., et al.\ 2010, \apj, 725, 2392 

\bibitem[2012]{mat12} Matthews, L.~D., Marengo, M., Evans, N.~R., \& Bono, G.\ 2012, \apj, 744, 53

\bibitem[2005]{mer05} M{\'e}rand, A., Kervella, P., Coud{\'e} du Foresto, V., et al.\ 2005, \aap, 438, L9 

\bibitem[2006]{mer06} M{\'e}rand, A., Kervella, P., Coud{\'e} du Foresto, V., et al.\ 2006, \aap, 453, 155

\bibitem[1984]{mof84} Moffett, T.~J., \& Barnes, T.~G., III 1984, \apjs, 55, 389 

\bibitem[1985]{mof85} Moffett, T.~J., \& Barnes, T.~G., III 1985, \apjs, 58, 843 

\bibitem[2004]{nar04} Nardetto, N., Fokin, A., Mourard, D., et al.\ 2004, \aap, 428, 131 

\bibitem[2006a]{nar06} Nardetto, N., Mourard, D., Kervella, P., et al.\ 2006a, \aap, 453, 309 

\bibitem[2006b]{nar06b} Nardetto, N., Fokin, A., Mourard, D., \& Mathias, P.\ 2006b, \aap, 454, 327 

\bibitem[2007]{nar07} Nardetto, N., Mourard, D., Mathias, P., Fokin, A., \& Gillet, D.\ 2007, \aap, 471, 661 

\bibitem[2009]{nar09} Nardetto, N., Gieren, W., Kervella, P., Fouqu{\'e}, P., Storm, J., Pietrzynski, G., Mourard, D., \& Queloz, D.\ 2009, \aap, 502, 951 

\bibitem[2011]{nar11} Nardetto, N., Fokin, A., Fouqu{\'e}, P., et al.\ 2011, \aap, 534, L16 

\bibitem[2010]{nei10} Neilson, H.~R., Ngeow, C.-C., Kanbur, S.~M., \& Lester, J.~B.\ 2010, \apj, 716, 1136 

\bibitem[2011]{nei11} Neilson, H.~R., \& Lester, J.~B.\ 2011, \aap, 530, A65 

\bibitem[2012]{nei11a} Neilson, H.~R., Nardetto, N., Ngeow, C.-C., Fouqu{\'e}, P. \& Storm, J.\ 2012, \aap, 541, A134

\bibitem[2005]{nge05} Ngeow, C.-C., Kanbur, S.~M., Nikolaev, S., et al.\ 2005, \mnras, 363, 831 

\bibitem[2009]{nge09} Ngeow, C.-C., Kanbur, S.~M., Neilson, H.~R., Nanthakumar, A., \& Buonaccorsi, J.\ 2009, \apj, 693, 691 

\bibitem[2004]{san04} Sandage, A., Tammann, G.~A., \& Reindl, B.\ 2004, \aap, 424, 43 

\bibitem[2004]{sto04} Storm, J., Carney, B.~W., Gieren, W.~P., et al.\ 2004, \aap, 415, 531 

\bibitem[2011]{sto11a} Storm, J., Gieren, W., Fouque, P., et al.\ 2011, \aap, 534, A94

\bibitem[2010]{tur10} Turner, D.~G.\ 2010, \apss, 326, 219 

\bibitem[2007]{van07} van Leeuwen, F., Feast, M.~W., Whitelock, P.~A., \& Laney, C.~D.\ 2007, \mnras, 379, 723 


\end{thebibliography}
\end{document}